\documentclass[12pt,a4paper]{article}
\usepackage{amsfonts,latexsym}
\usepackage{graphicx,color}
\usepackage{dcolumn}
\usepackage{graphicx}
\usepackage{amsmath}
\usepackage{amsfonts}
\usepackage{amssymb}

\renewcommand{\thefootnote}{\fnsymbol{footnote}}

\begin{document}

\vspace{12mm}

\begin{center}
{{{\Large {\bf Rotating scalarized black holes in  scalar couplings to two topological terms}}}}\\[10mm]

{De-Cheng Zou$^{a}$\footnote{e-mail address: dczou@yzu.edu.cn} and   Yun Soo Myung$^b$\footnote{e-mail address: ysmyung@inje.ac.kr}}\\[8mm]

{${}^a$Center for Gravitation and Cosmology and College of Physical Science and Technology, Yangzhou University, Yangzhou 225009, China\\[0pt]}
{${}^b$Institute of Basic Sciences and Department  of Computer Simulation, Inje University Gimhae 50834, Korea\\[0pt] }

\end{center}
\vspace{2mm}

\begin{abstract}
The tachyonic instability  of the Kerr  black holes is analyzed
in the  Einstein-scalar theory with the quadratic scalar couplings  to two topological terms which  are parity-even Gauss-Bonnet and parity-odd Chern-Simons terms. For  positive coupling $\alpha$, we use the (2+1)-dimensional  hyperboloidal  foliation method to derive the threshold curve which is the boundary between stable and unstable Kerr black holes by considering a spherically symmetric  scalar-mode
perturbation.  In  case of  negative coupling,   a newly bound of   $a\ge0.26$ with $a$ the rotation parameter is found for unstable Kerr black holes and its threshold curve is derived.

\end{abstract}
\vspace{5mm}

\vspace{1.5cm}

\hspace{11.5cm}
\newpage
\renewcommand{\thefootnote}{\arabic{footnote}}
\setcounter{footnote}{0}


\section{Introduction}
Black hole scalarization is  well understood as a good  mechanism to have a black hole with scalar hair.
When a scalar field couples to either Gauss-Bonnet term~\cite{Doneva:2017bvd,Silva:2017uqg,Antoniou:2017acq} or to Maxwell term~\cite{Herdeiro:2018wub}, the  tachyonic instability for general relativity (GR) black holes represents the hallmark of the spontaneous scalarization. As the linear  instability develops and the scalar field  grows, nonlinear terms play the important role
and quench the instability. Eventually, the scalarzied black hole solutions are constructed numerically as infinite $n=0,~1,~2,~\cdots$ black holes where
the fundamental ($n=0$) scalarized  black hole is stable against radial perturbations whereas  radially  excited  black holes with $n>0$ turn out to be unstable, irrespective of coupling terms~\cite{Zou:2020zxq}. This implies that  the fundamental scalarized   black hole is considered as an endpoint of the GR black hole.

Now, we wish to mention briefly  the onset of scalarization for rotating GR black holes.
Spontaneous scalarization of Kerr black holes has been firstly studied in a scalar coupling to the Gauss-Bonnet term  with positive coupling~\cite{Cunha:2019dwb,Collodel:2019kkx}, implying that  the high rotation with $a\ge 0.5$ suppresses scalarization.
Recently, an $a$-bound of $a/M \ge 0.5$ (high rotation) was found as the onset of scalarization for  Kerr black holes  with  negative coupling~\cite{Dima:2020yac}.
It implies that there is a minimum rotation $a_{\rm min}/M=0.5$ below which the instability never appears.  This $a$-bound  was recovered  analytically  in~\cite{Hod:2020jjy} and numerically in~\cite{Doneva:2020nbb}.
In this direction,  the spin-induced scalarized  black holes were numerically constructed for the high rotation  and negative coupling~\cite{Herdeiro:2020wei,Berti:2020kgk}.

On the other hand, it is found that no such an $a$-bound exists when investigating  the tachyonic instability of Kerr black holes in a scalar coupling to the Chern-Simons term with negative coupling~\cite{Myung:2020etf}.  This suggests  that the odd-parity Chern-Simons term plays a different role  from the even-parity Gauss-Bonnet term.

In this work, we wish to perform the instability analysis of the Kerr  black holes
in  scalar couplings to two topological terms  with the  same quadratic coupling parameter $\alpha$.
The organization of our work is as follows.
In section 2, we mention Kerr black holes without scalar hair in the Einstein-scalar-Gauss-Bonnet-Chern-Simons (EsGBCS) theory.
We discuss the tachyonic instability of Kerr black holes for both positive and negative $\alpha$ in section 3.
It is essential to  derive the time evolution of a spherically symmetric scalar mode ($l=m=0$) for the instability analysis because the Kerr background is a stationary, axisymmetric, and non-static spacetime.

We will use  the (2+1)-dimensional  hyperboloidal foliation method to derive the time evolution of a spherically symmetric scalar-mode.
It will take a long time to complete the computations even though we confine ourselves to  a spherically symmetric scalar-mode.
 For positive $\alpha$, we wish to  derive the threshold curve which is the boundary between stable and unstable Kerr black holes.   Curiously, we expect  to obtain a different  $a$-bound of $a\ge 0.26$, in addition to  the threshold curve,  for negative $\alpha$.
In section 4, we will discuss our results by comparing with those of slowly rotating black holes in the EsGBCS theory.

\section{Kerr black holes }

Here, we consider the action of  Einstein-scalar-Gauss-Bonnet-Chern-Simons (EsGBCS) theory~\cite{Myung:2021fzo}  as
\begin{eqnarray}
S_{\rm EsGBCS}=\frac{1}{16 \pi}\int d^4 x\sqrt{-g} \Big[
R-\frac{1}{2}(\partial \phi)^2+f(\phi)(R^2_{\rm GB}+{}^{*}RR)\Big],\label{Action}
\end{eqnarray}
where we use geometric units of $G=c=1$. $\phi$ is a real scalar field and $f(\phi)$ is the  coupling function coupled  to two topological terms: the  Gauss-Bonnet term
$ R^2_{\rm GB}=R^2-4R_{\mu\nu}R^{\mu\nu}+R_{\mu\nu\rho\sigma}R^{\mu\nu\rho\sigma}$ and  the
Chern-Simons term
${}^{*}RR=\frac{1}{2}\epsilon^{\mu\nu\rho\sigma}R^{\eta}_{~\xi\rho\sigma}R^\xi_{~\eta\mu\nu}$.
Here we choose  the  quadratic coupling function of $f(\phi)=\alpha \phi^2$ because it provides the simplicity for spontaneous scalarization.
To allow for spontaneous scalarization, the coupling
function $f(\phi)$ should possess certain properties. The GR (Kerr) black hole solutions should remain solutions of
the theory. This is the case when the Gauss-Bonnet and  Chern-Simons terms do not contribute to the field equations.
Selecting a coupling function $f(\phi)$ such that $f'(\phi)=0$ for $\phi=0$, the source term in the scalar equation of $\nabla^2\phi+f'(\phi) (R^2_{\rm GB}+{}^{*}RR)=0$ vanishes for $\phi = 0$. It  implies  that  $\phi = 0$ is a solution. This is why we choose a non-minimally coupling function $f(\phi)$. In this direction,  a linear function of $f(\phi)=\alpha \phi$ is excluded. Also, if one chooses no scalar coupling like $f(\phi)=\alpha$, the Gauss-Bonnet and  Chern-Simons terms do not contribute to the field equations absolutely because they are topological terms.

Firstly, we derive the Einstein  equation
\begin{eqnarray}
&&G_{\mu\nu}=\frac{1}{2}\partial_{\mu}\phi\partial_{\nu}\phi- \frac{1}{4}g_{\mu\nu}(\partial \phi)^2-\alpha \phi^2 H_{\mu\nu}-4\alpha [\nabla^\rho \nabla^\sigma (\phi^2)P_{\mu\rho\nu\sigma}+ C_{\mu\nu}],\label{eqn-1}
\end{eqnarray}
where $H_{\mu\nu}$ and
$P_{\mu\rho\nu\sigma}$ take the forms
\begin{eqnarray}
H_{\mu\nu}&=&2(RR_{\mu\nu}-2R_{\mu\rho}R^\rho~_\nu -2R^{\rho\sigma}R_{\mu\rho\nu\sigma}+R_\mu~^{\rho\sigma\lambda}R_{\nu\rho\sigma\lambda})-\frac{1}{2}g_{\mu\nu}R^2_{\rm GB}, \label{H-ten}\\
P_{\mu\rho\nu\sigma}&=&R_{\mu\rho\nu\sigma}+g_{\mu\sigma}R_{\nu\rho}-g_{\mu\nu}R_{\rho\sigma}+g_{\nu\rho}R_{\mu\sigma}-g_{\rho\sigma}R_{\mu\nu}
+\frac{R}{2}(g_{\mu\nu}g_{\rho\sigma}-g_{\mu\sigma}g_{\nu\rho}). \label{p-ten}
\end{eqnarray}
If $f(\phi)=\alpha$, $P_{\mu\nu\rho\sigma}$-term disappears, implying no modifications of the Einstein equation.
On the other hand,  the Cotton tensor  $C_{\mu\nu}$ is given by
\begin{eqnarray}\label{cotton}
C_{\mu\nu}=\nabla_{\rho}(\phi^2)~\epsilon^{\rho\sigma
\gamma}_{~~~~(\mu}\nabla_{\gamma}R_{\nu)\sigma}+\frac{1}{2}\nabla_{\rho}\nabla_{\sigma}
(\phi^2)~\epsilon_{(\nu}^{~~\rho \gamma
\delta}R^{\sigma}_{~~\mu)\gamma \delta}
\end{eqnarray}
with $\epsilon_{\mu\nu\rho\sigma}$  being the 4-dimensional Levi-Civita tensor.
However, an important scalar equation  is found to be
\begin{equation}
\nabla^2\phi+2\alpha (R^2_{\rm GB}+{}^{*}RR)\phi=0.\label{eqn-2}
\end{equation}

 Even though the Einstein equation (\ref{eqn-1}) looks like a complicated form, it is easy to find Kerr black holes when choosing a trivial scalar field.
 Actually, Eq. (\ref{eqn-1}) together with $\bar{\phi}=0$ reduces to $\bar{R}_{\mu\nu}=0$ which implies   the rotating GR black hole (Kerr spacetime) written in terms of the Boyer-Lindquist coordinates
\begin{eqnarray}
ds_{\rm Kerr}^2&=&\bar{g}_{\mu\nu}dx^{\mu}dx^{\nu} \nonumber \\
&=& -\frac{\Delta}{\rho^2}(dt-a \sin^2 \theta d\varphi)^2+\frac{\rho^2}{\Delta}dr^2
+\rho^2 d\theta^2 +\frac{\sin^2 \theta}{\rho^2}[a dt -(r^2+a^2)d\varphi]^2,  \label{Kerr-sp}
\end{eqnarray}
where
\begin{equation}
\Delta=r^2-2Mr +a^2,~\rho^2=r^2+a^2 \cos^2\theta \label{delta-rho}
\end{equation}
with  mass $M$ and  rotation parameter $a=J/M>0$.
We mention that Eq. (\ref{Kerr-sp}) describes a stationary, axisymmetric and non-static  spacetime because it does not depend on time, it does not depend on $\phi$, and it is not invariant under $t\to -t$.
From  $\Delta=0$, one finds outer/inner horizons
\begin{equation}
r_\pm=M\Big[1\pm \sqrt{1-a^2/M^2}\Big],
\end{equation}
while one gets the angular velocity of rotating black hole as $\Omega_{\rm H}=a/(r_+^2+a^2)$.

\section{Tachyonic instability of Kerr black holes}
We wish to study whether there is a regime within which the Kerr black hole solution is unstable against perturbations in the EsGBCS theory.
To this end, we have to obtain  the linearized theory by linearizing the Einstein and scalar equations.
Introducing two perturbations ($h_{\mu\nu},\delta \phi$) around the Kerr background
\begin{eqnarray} \label{m-p}
g_{\mu\nu}=\bar{g}_{\mu\nu}+h_{\mu\nu},~~\phi=0+\delta\phi,
\end{eqnarray}
the linearized equation to Eq. (\ref{eqn-1}) takes a simple form
\begin{eqnarray}\label{pertg}
\delta R_{\mu\nu}(h)=0\to \bar{\nabla}^{\gamma}\bar{\nabla}_{\mu}
h_{\nu\gamma}+\bar{\nabla}^{\gamma}\bar{\nabla}_{\nu}
h_{\mu\gamma}-\bar{\nabla}^2h_{\mu\nu}-\bar{\nabla}_{\mu} \bar{\nabla}_{\nu} h=0.
\end{eqnarray}
The tensor-stability analysis for the Kerr black hole with Eq. (\ref{pertg}) is the same as in the general relativity.
It is found that there are  no exponentially growing tensor modes propagating around the Kerr  background by making use of the null tetrad formalism~\cite{chandra}.
Before we proceed, let us consider the Klein-Gordon equation~\cite{Myung:2013oca}
\begin{eqnarray}
(\bar{\nabla}^2-\mu^2)\delta\phi=0\label{phi-KG}
\end{eqnarray}
with $\mu^2$  mass squared. This is
is a linearized  equation for the massive scalar propagating  around the Kerr background.
Reminding the axisymmetric background shown in Eq. (\ref{Kerr-sp}), it is convenient to separate the scalar
into modes
\begin{equation}
\delta \phi(t,r,\theta,\phi)=\sum_{lm}e^{-i\omega t + i m \phi} S_{\ell m
}(\theta){\cal R}_{\ell m}(r)\,, \label{4-sep}
\end{equation}
where $S_{\ell m}(\theta)$ denotes  spheroidal harmonics with $-m\le \ell
\le m$ and ${\cal R}_{\ell m}(r)$  represents radial function. We may choose a positive frequency $\omega$ of
the mode.  Plugging Eq. (\ref{4-sep}) into Eq. (\ref{phi-KG}), the
angular and radial (Teukolsky) equations for $S_{\ell m}(\theta)$ and ${\cal
R}_{\ell m}(r)$ are given by
\begin{eqnarray}
 \frac{1}{\sin \theta}\partial_{\theta}\left (
\sin \theta
\partial_{\theta} S_{\ell m} \right )&+& \left [  a^2 (\omega^2-\mu^2) \cos^2
\theta-\frac{m^2}{\sin ^2{\theta}}+\lambda_{lm} \right ]S_{\ell m} =0, \label{waveeq-S}\\
 \Delta \partial_r \left ( \Delta \partial_r {\cal
R}_{\ell m}(r) \right )&-& [\Delta U-K^2] {\cal R}_{\ell m}(r)=0 \label{waveeq-P}
\end{eqnarray}
with $U=\mu^2(r^2+a^2)-2am\omega +\lambda_{lm}$ and $K=\omega(r^2+a^2)-am$.
Here $\lambda_{lm}$ is the separation constant  given
by
\begin{eqnarray}
\lambda_{lm}=l(l+1)+\sum^\infty_{k=1}c_ka^{2k}(\mu^2-\omega^2)^k
 \label{eigenvalues}
\end{eqnarray}
for $\omega \simeq M$ only. Introducing the tortoise  coordinate $r_*$  defined    by $dr_*=
\frac{r^2+a^2}{\Delta}dr$ and $\psi(r)=\sqrt{r^2+a^2}{\cal R}(r)$,  the Teukolsky equation takes the
Schr\"odinger form
\begin{eqnarray}
\frac{d^2\psi}{dr_*^2}+\Big[\omega^2-V(r,\omega)\Big]\psi=0\\ \label{sch-eq}
\end{eqnarray}
with the potential
\begin{eqnarray}
V(r,\omega)=\omega^2&-&\frac{3\Delta^2r^2}{(a^2+r^2)^4}+\frac{\Delta[\Delta+2r(r-M)]}{(a^2+r^2)^3} \nonumber \\
&+&\frac{\Delta \mu^2}{a^2+r^2}-\Big[\omega-\frac{am}{a^2+r^2}\Big]^2 \nonumber \\
&-&\frac{\Delta}{(a^2+r^2)^2}\Big[2am\omega +a^2(\mu^2-\omega^2)-\lambda_{lm}\Big]. \label{c-pot}
\end{eqnarray}
This potential  is chosen for studying the superradiant instability~\cite{Zouros:1979iw}.
In order to find if there exists a trapping potential (a necessary condition for superradiant instability),
one should analyze the shape of potential $V(r,\omega)$ carefully.

Now let us go back to the linearized scalar theory in the EsGBCS theory.
The instability of Kerr black holes will be determined by the linearized scalar equation whose form  is given by
\begin{eqnarray}
 \Big(\bar{\nabla}^2-\mu^2_{\rm eff}\Big)\delta\phi=0\label{phi-eq2}
\end{eqnarray}
with  an effective mass
\begin{equation} \label{eff-m}
\mu^2_{\rm eff}\equiv\mu^2_{\rm GB}+\mu^2_{\rm CS}=-2\alpha \bar{R}^2_{\rm GB}-2\alpha~{}^{*}\bar{R}\bar{R}.
\end{equation}
Here, we have
\begin{eqnarray}
&&\bar{R}^2_{\rm GB}=\frac{48M^2(r^6-15r^4 a^2\cos^2\theta+15 r^2a^4\cos^4\theta-a^6\cos^6\theta)}{(r^2+a^2\cos^2\theta)^6} \label{GB1} \\
&&\quad\quad \simeq \frac{48M^2}{r^6}\Big(1-\frac{21a^2\cos^2\theta}{r^2}+\cdots\Big)  \label{GB2}
\end{eqnarray}
and
\begin{eqnarray}
&&{}^{*}\bar{R}\bar{R}=\frac{96rM^2a\cos \theta(3r^4-10r^2a^2\cos^2\theta+3a^4\cos^4\theta)}{(r^2+a^2\cos^2\theta)^6} \label{CS1} \\
  &&\quad\quad \simeq  \frac{288M^2 a \cos\theta}{r^7} \Big(1-\frac{28a^2\cos^2\theta}{3r^2}+\cdots\Big).             \label{CS2}
\end{eqnarray}
Under the parity transformation, one finds that even: $\bar{R}^2_{\rm GB}(\pi-\theta)\to \bar{R}^2_{\rm GB}(\theta)$ and odd: ${}^{*}\bar{R}\bar{R}(\pi-\theta)\to -{}^{*}\bar{R}\bar{R}(\theta)$.
At this stage, we mention  that  Eqs. (\ref{GB2}) and (\ref{CS2}) represent series forms written in terms of $a$.
In the slowly rotating black holes with $a\ll1$, one  considers the first terms in Eqs. (\ref{GB2}) and (\ref{CS2}) only~\cite{Myung:2021fzo}.
It is important to note  that $\mu^2_{\rm GB}(\mu^2_{\rm CS})$
is odd (even)  under a combined transformation of $\alpha\to -\alpha$ and $\theta\to\pi-\theta$.
This implies that  positive and negative $\alpha$  will show different results.
From now on, we consider two cases of $\alpha>0$ and $\alpha<0$ separately.

\begin{figure*}[t!]
   \centering
  \includegraphics{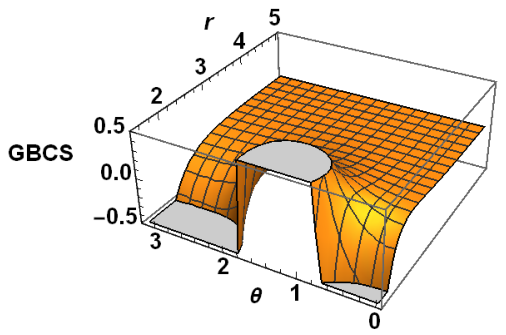}
  \hfill%
  \includegraphics{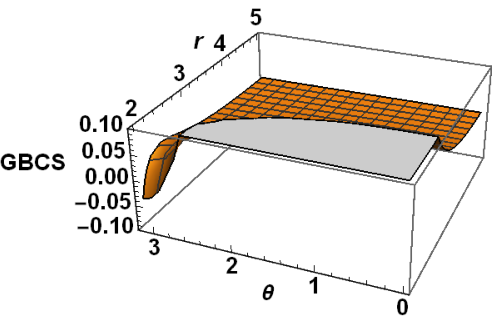}
  \hfill%
  \includegraphics{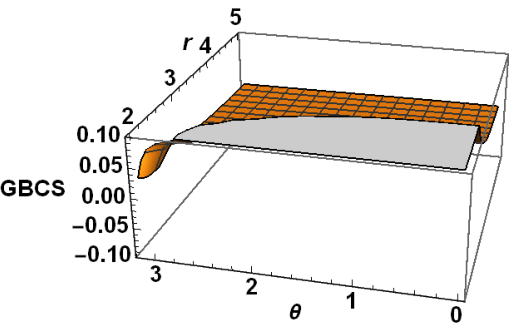}
\caption{ Three 3D graphs for GBCS($=\bar{R}^2_{\rm GB}+{}^{*}\bar{R}\bar{R}$) with $M=1$. These include $r\in[r_+,5]$ and $\theta\in[0,\pi]$. Left: a graph for $a=0.87$ indicating  positive region with  negative regions around the poles of $\theta=0,\pi$. Middle: a graph for $a=0.27>0.26$ representing  positive region with a tiny negative region   at $\theta=\pi$.
Right: a  graph for $a=0.25<0.26$ showing whole positive region. }
\end{figure*}

\subsection{Positive coupling $\alpha$}

First of all, we note that  $\bar{R}^2_{\rm GB}$  is an  even function, while  ${}^{*}\bar{R}\bar{R}$  is an odd function.
For  $\bar{R}^2_{\rm GB}+{}^{*}\bar{R}\bar{R}(\mu^2_{\rm eff}$ in Eq.(\ref{eff-m})), however, one observes from Fig. 1 that negative (positive) region appears for $a\ge0.26$. For $a< 0.26$, $\bar{R}^2_{\rm GB}+{}^{*}\bar{R}\bar{R}(\mu^2_{\rm eff}$ in Eq.(\ref{eff-m})) becomes positive (negative) and thus,  enhancing scalarization.
Our observation comes from $\bar{R}^2_{\rm GB}$ and ${}^{*}\bar{R}\bar{R}$ solely but not from including the  coupling parameter $\alpha$.
It is well known that the threshold  curve $\alpha=\alpha_{\rm th}(a)$ for Kerr black holes depends on  $a$.  However,  its form will be determined by performing  numerical computations for a long time.

The separation of scalar given by Eq. (\ref{4-sep}) is impossible to occur here because of $\mu^2_{\rm eff}(r,\theta,\alpha)$.
Instead, we introduce  the separation of variables after transforming to the ingoing Kerr-Schild coordinates $\{\tilde{t},r,\theta,\tilde{\varphi}\}$
 \begin{equation}
 \delta \phi(\tilde{t},r,\theta,\tilde{\varphi}) =\frac{1}{r} \sum_{m} u_m(\tilde{t},r,\theta) e^{i m \tilde{\varphi}}. \label{1-sep}
\end{equation}
Plugging Eq. (\ref{1-sep}) into  Eq. (\ref{phi-eq2}) leads to the (2+1)-dimensional Teukolsky equation as
\begin{equation}
A^{\tilde{t}\tilde{t}}\partial_{\tilde{t}}^2u_m+A^{\tilde{t}r}\partial_{\tilde{t}}\partial_ru_m+A^{rr}\partial^2_r u_m+
A^{\theta\theta}\partial_\theta^2u_m+B^{\tilde{t}}\partial_{\tilde{t}}u_m
+B^r\partial_r u_m+B^\theta\partial_\theta u_m+C u_m=0\label{phi-eq4}
\end{equation}
with coefficients
\begin{eqnarray}
&&A^{\tilde{t}\tilde{t}}=\rho^2+2Mr,~~A^{\tilde{t}r}=-4Mr,~~A^{rr}=-\Delta,~~A^{\theta\theta}=-1,\nonumber\\
&&B^{\tilde{t}}=2M,~~B^r=\frac{2}{r}(a^2-Mr)-2ima,~~B^\theta=-\cot\theta,  \label{coeffs}\\
&&C=\frac{m^2}{\sin^2\theta}-\frac{2(a^2-Mr)}{r^2}+\frac{2ima}{r}+\mu^2_{\rm eff}\rho^2. \nonumber
\end{eqnarray}
We employ the (2+1)-dimensional hyperboloidal foliation method by introducing the compactified radial coordinates $R$ and suitable time coordinates $\tau$~\cite{Racz:2011qu} to solve Eq. (\ref{phi-eq4}) numerically for the CS term~\cite{Gao:2018acg} and the GB term~\cite{Zhang:2020pko}.
Here, it is not necessary  to describe this method explicitly because we have already used it to compute  the time evolution of a spherically symmetric scalar-mode~\cite{Myung:2020etf,Myung:2021fzo}.
The differential equations in $R$ and $\theta$ are solved by using  the finite difference method and the time ($\tau$) evolution is obtained by adopting
the fourth-order Runge-Kutta integrator when computing $u_{lm}(\tau,R,\theta)$.
\begin{figure*}[t!]
   \centering
  \includegraphics{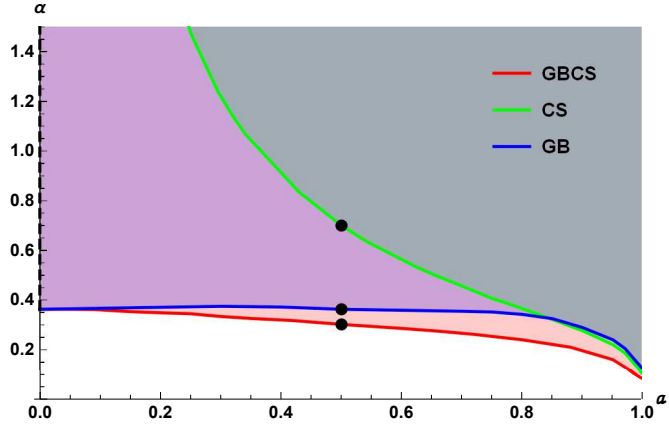}
\caption{Three threshold curves [$\alpha=\alpha_{\rm th}(a)$] being the boundary between stable  and unstable Kerr black holes  by observing time evolution of a $l=m=0$-scalar mode  for positive  $\alpha$.
The GBCS, CS, and GB  represent the threshold curves for the presence of  Gauss-Bonnet-Chern-Simons, Chern-Simons, and  Gauss-Bonnet terms, respectively.
Dots ($\bullet$) represent  three different points of $\alpha_{\rm th}=$0.7(CS), 0.363(GB), and 0.302(GBCS) at  $a=0.5$
Dashed line ($\alpha \ge 0.363$) on the $\alpha$-axis is designed for unstable Schwarzschild black holes in the non-rotating limit of $a\to 0$.
}
\end{figure*}
As an initial scalar mode, we may introduce   a Gaussian function [$u_{lm}(\tau=0,R,\theta)\sim Y_{lm}(\theta)e^{-\frac{(R-R_c)^2}{2\sigma^2}}$] with spherical harmonics $Y_{lm}(\theta)$ localized at $R=R_c$ outside the outer horizon.
Since the Kerr spacetime is axisymmetric,  the mode coupling may occur such that
a purely even (odd) initial multipole $l$  will excite other even (odd) multipoles (denoted by $l'$) with the same $m$ as it evolves. So, one may consider $l=0$ and $l = 1$  as representative even and odd multipoles with  axisymmetric perturbations with $m = 0$ for the GB term~\cite{Zhang:2020pko}. For the CS term~\cite{Gao:2018acg}, the axisymmetric ($m=0$) and non-axisymmetric ($m\not=0$)  cases were considered.
Here, however,  we consider a spherically symmetric scalar-mode of $l=m=0$ only because it needs  much  computation times to carry out six different cases: for $\alpha>0$, $\alpha^{\rm GB}_{\rm th}(a),~\alpha^{\rm CS}_{\rm th}(a),~\alpha^{\rm GBCS}_{\rm th}(a)$ and for $\alpha<0$, $-\alpha^{\rm GB}_{\rm th}(a),~-\alpha^{\rm CS}_{\rm th}(a),~-\alpha^{\rm GBCS}_{\rm th}(a)$. Also, the spherically symmetric scalar-mode  could be considered as  a representative for all scalar modes when performing tachyonic instability analysis of scalar modes around the Kerr black hole~\cite{Dima:2020yac}.

From Fig. 2, we find  three threshold curves (existence curves) $\alpha_{\rm th}(a)$  which are the boundary between  stable and unstable regions
based on the time evolutions of a scalar mode $u_{00}$. We note the range of $\alpha_{\rm th}$: $\alpha^{\rm GB}_{\rm th},~\alpha^{\rm GBCS}_{\rm th}\in(0,0.363]$ but $\alpha^{\rm CS}_{\rm th}\in(0,\infty)$.
The CS-threshold curve decreases rapidly as $a$ increases, while it never hits the $\alpha$-axis in the non-rotating  limit of $a\to0$~\cite{Myung:2020etf}.
On the other hand, the GB-and GBCS-threshold curves start  at $\alpha=0.363$  on the $\alpha$-axis corresponding to the threshold of unstable Schwarzschild black hole,
whereas they   decrease  slowly as $a$ increases~\cite{Collodel:2019kkx}.
The unshaded region [$\alpha<\alpha_{\rm th}(a)$: no growing mode] of each curve represents the stable Kerr black holes, while the shaded region [$\alpha>\alpha_{\rm th}(a)$:  growing mode] denotes the unstable Kerr black holes. We call the latter as `$a$ dependent $\alpha$-bound' for onset of rotating scalarization.
Also, Fig. 2 includes  the stable and unstable (dashed-line: $\alpha\ge0.363$) Schwarzschild black holes on the $\alpha$-axis.
We observe that the unstable region increases as $\alpha_{\rm th}$ decreases for fixed $a=0.5$: the largest $\alpha_{\rm th}=0.7$ is obtained for the CS term, the medium $\alpha_{\rm th}=0.363$ is for the GB term, and the smallest $\alpha_{\rm th}=0.302$ is for the GBCS term. This could be read off from the fact
that the role of $a$ is critical in Eq. (\ref{CS2}),  but it is less  critical in Eq. (\ref{GB2}).
It is worth noting  that  we have  $\alpha_{\rm th}$=0.086 (GBCS), 0.107(CS), and 0.126(GB) in the nearly extremal limit of $a=0.998$.

Concerning  the reliability of numerical tests for positive $\alpha$,  our data includes $(a,\alpha)=\{(0.998,0.107),(0.362,1.0)\}$ for CS case appeared in ~\cite{Gao:2018acg} and $(a,\alpha)=\{(0,0.363),(0.3,0.375),\\(0.5,0.363),(0.9,0.3)\}$ for GB case (when replacing $\alpha$ by $2\alpha$)  appeared in~\cite{Zhang:2020pko}. Here, we note that 
the precision and accuracy of the (2+1)-dimensional hyperboloidal foliation method was tested in Refs.\cite{Gao:2018acg,Zhang:2020pko}.
 Furthermore, our data of  GBCS for $0\le a\le 0.1$ is the nearly same as obtained  from a spherically symmetric scalar mode propagating around slowly rotating black holes~\cite{Myung:2021fzo}.
Finally, we recover $\alpha_{\rm th}=0.363$ for Schwarzschild black hole from GB and GBCS cases when choosing $a=0$.
\subsection{Negative coupling $\alpha$ }
We observe  from Fig. 1 that a necessary condition for  $\bar{R}^2_{\rm GB}+{}^{*}\bar{R}\bar{R}$($\mu^2_{\rm eff}$ in Eq.(\ref{eff-m})) to have some negative region is an $a$-bound of  $a\ge0.26$, enhancing  scalarization. The other case of $a<0.26$ is not allowed for tachyonic  instability.
\begin{figure*}[t!]
   \centering
  \includegraphics{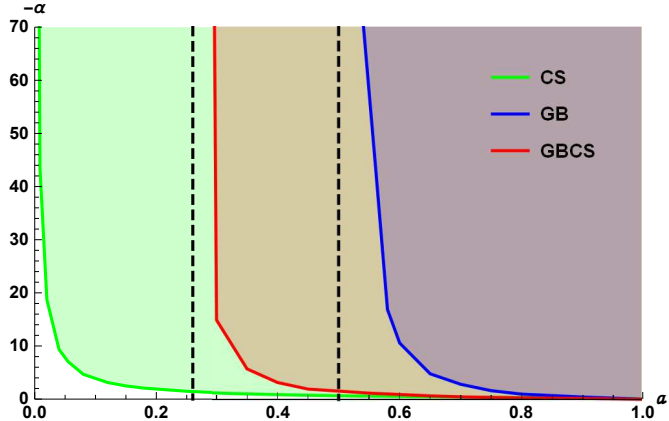}
\caption{Three threshold curves [$-\alpha=-\alpha_{\rm th}(a)$] being the boundary between stable  and unstable Kerr black holes  by observing time evolution of $l=m=0$-scalar mode  for negative $\alpha$.
Two dashed lines represent $a=0.26$ and $a=0.5$ as  minimum values for the $a$-bound in the GBCS and GB cases, respectively. }
\end{figure*}
From Fig. 3, we confirm  that the $a$-bound for the EsGBCS theory is $a\ge 0.26$, while the $a$-bound for the EsGB theory denotes $a\ge 0.5$. This implies that an addition  of the CS term to the GB term has shifted from  the bound of $a\ge 0.5$ to $a\ge 0.26$. We remind the reader that the GBCS and GB are not invariant  under the combined transformation of $\theta\to \pi-\theta$ and $\alpha\to -\alpha$. So, we  have different threshold curves for negative $\alpha$ (Fig. 2)  when comparing with positive $\alpha$ (Fig. 3).
On the other hand, there is no $a$-bound for the CS term with negative $\alpha$.
This curve is the same as in Fig. 2 because $\mu^2_{\rm CS}$ is invariant under the combined transformation.

Finally, concerning  the reliability of numerical tests for negative $\alpha$,  our data includes $(a,-\alpha)=\{(0.7,2.9),(0.8,1.0),(0.9,0.46)\}$ for GB case (when replacing $-\alpha$ by $-2\alpha$)  appeared in~\cite{Zhang:2020pko}.

\section{Discussions }
First of all, we mention the tachyonic instability for  slowly rotating black holes in the EsGBCS theory.
It was shown  that
slowly rotating black holes with $a\ll 1$  are unstable against a spherically symmetric  scalar-mode of $l=m=0$
for  positive coupling $\alpha$ only~\cite{Myung:2021fzo}.  A threshold  curve   $\alpha=\alpha_{\rm th}(a)$  which is the boundary between  stable and unstable black holes was derived  by considering  the constant scalar modes under time evolution.  For negative coupling,  there is  no tachyonic instability for scalarization since the mass term of $\mu^2_{\rm eff}$ is always  positive outside the outer  horizon. Therefore,    any  $a$-bound is not found.

In this work, we have investigated  the tachyonic instability for Kerr black holes in the same theory, allowing whole range of $0\le a \le 1$.
Since $\mu^2_{\rm GB}(\mu^2_{\rm CS})$
is variant (invariant)  under a combined transformation of $\alpha\to -\alpha$ and $\theta\to\pi-\theta$,
positive and negative $\alpha$  have shown different results. For positive $\alpha$, we have obtained the threshold curve for the GBCS case (see Fig. 2) which is the nearly same as that for slowly rotating black holes for sufficiently low rotation of $a\ll 1$. On the other hand, for negative $\alpha$, we have obtained an $a$-bound of $a\ge 0.26$
and the threshold curve for the GBCS case (see Fig. 3), which is never found from the instability analysis for  slowly rotating black holes.

 \vspace{1cm}

{\bf Acknowledgments}

 This work was supported by the National Research Foundation of Korea (NRF) grant funded by the Korea government (MOE)
 (No. NRF-2017R1A2B4002057).
 

\vspace{1cm}


\newpage
\begin{thebibliography}{99}

\bibitem{Doneva:2017bvd}
  D.~D.~Doneva and S.~S.~Yazadjiev,
  Phys.\ Rev.\ Lett.\  {\bf 120}, no. 13, 131103 (2018)
  doi:10.1103/PhysRevLett.120.131103
  [arXiv:1711.01187 [gr-qc]].

\bibitem{Silva:2017uqg}
  H.~O.~Silva, J.~Sakstein, L.~Gualtieri, T.~P.~Sotiriou and E.~Berti,
  Phys.\ Rev.\ Lett.\  {\bf 120}, no. 13, 131104 (2018)
  doi:10.1103/PhysRevLett.120.131104
  [arXiv:1711.02080 [gr-qc]].

\bibitem{Antoniou:2017acq}
  G.~Antoniou, A.~Bakopoulos and P.~Kanti,
  Phys.\ Rev.\ Lett.\  {\bf 120}, no. 13, 131102 (2018)
  doi:10.1103/PhysRevLett.120.131102
  [arXiv:1711.03390 [hep-th]].

\bibitem{Herdeiro:2018wub}
  C.~A.~R.~Herdeiro, E.~Radu, N.~Sanchis-Gual and J.~A.~Font,
  Phys.\ Rev.\ Lett.\  {\bf 121}, no. 10, 101102 (2018)
  doi:10.1103/PhysRevLett.121.101102
  [arXiv:1806.05190 [gr-qc]].


\bibitem{Zou:2020zxq}
  D.~C.~Zou and Y.~S.~Myung,
  Phys.\ Rev.\ D {\bf 102}, no. 6, 064011 (2020)
  doi:10.1103/PhysRevD.102.064011
  [arXiv:2005.06677 [gr-qc]].


\bibitem{Cunha:2019dwb}
  P.~V.~P.~Cunha, C.~A.~R.~Herdeiro and E.~Radu,
  Phys.\ Rev.\ Lett.\  {\bf 123}, no. 1, 011101 (2019)
  doi:10.1103/PhysRevLett.123.011101
  [arXiv:1904.09997 [gr-qc]].

\bibitem{Collodel:2019kkx}
  L.~G.~Collodel, B.~Kleihaus, J.~Kunz and E.~Berti,
  Class.\ Quant.\ Grav.\  {\bf 37}, no. 7, 075018 (2020)
  doi:10.1088/1361-6382/ab74f9
  [arXiv:1912.05382 [gr-qc]].

\bibitem{Dima:2020yac}
  A.~Dima, E.~Barausse, N.~Franchini and T.~P.~Sotiriou,
  Phys.\ Rev.\ Lett.\  {\bf 125}, no. 23, 231101 (2020)
  doi:10.1103/PhysRevLett.125.231101
  [arXiv:2006.03095 [gr-qc]].


\bibitem{Hod:2020jjy}
  S.~Hod,
  Phys.\ Rev.\ D {\bf 102}, no. 8, 084060 (2020)
  doi:10.1103/PhysRevD.102.084060
  [arXiv:2006.09399 [gr-qc]].

\bibitem{Doneva:2020nbb}
  D.~D.~Doneva, L.~G.~Collodel, C.~J.~Krüger and S.~S.~Yazadjiev,
  Phys.\ Rev.\ D {\bf 102}, no. 10, 104027 (2020)
  doi:10.1103/PhysRevD.102.104027
  [arXiv:2008.07391 [gr-qc]].


\bibitem{Herdeiro:2020wei}
  C.~A.~R.~Herdeiro, E.~Radu, H.~O.~Silva, T.~P.~Sotiriou and N.~Yunes,
  Phys.\ Rev.\ Lett.\  {\bf 126}, no. 1, 011103 (2021)
  doi:10.1103/PhysRevLett.126.011103
  [arXiv:2009.03904 [gr-qc]].



\bibitem{Berti:2020kgk}
  E.~Berti, L.~G.~Collodel, B.~Kleihaus and J.~Kunz,
  Phys.\ Rev.\ Lett.\  {\bf 126}, no. 1, 011104 (2021)
  doi:10.1103/PhysRevLett.126.011104
  [arXiv:2009.03905 [gr-qc]].




\bibitem{Myung:2020etf}
  Y.~S.~Myung and D.~C.~Zou,
  Phys.\ Lett.\ B {\bf 814}, 136081 (2021)
  doi:10.1016/j.physletb.2021.136081
  [arXiv:2012.02375 [gr-qc]].



\bibitem{Myung:2021fzo}
  Y.~S.~Myung and D.~C.~Zou,
  arXiv:2103.06449 [gr-qc].

\bibitem{chandra}
S. Chandrasekhar,
 {\it The Mathematical Theory of Black Holes} (Oxford University Press, New York, 1983).

\bibitem{Myung:2013oca}
  Y.~S.~Myung,
  Phys.\ Rev.\ D {\bf 88}, no. 10, 104017 (2013)
  doi:10.1103/PhysRevD.88.104017
  [arXiv:1309.3346 [gr-qc]].



\bibitem{Zouros:1979iw}
  T.~J.~M.~Zouros and D.~M.~Eardley,
  Annals Phys.\  {\bf 118}, 139 (1979).
  doi:10.1016/0003-4916(79)90237-9


\bibitem{Racz:2011qu}
  I.~Racz and G.~Z.~Toth,
  Class.\ Quant.\ Grav.\  {\bf 28}, 195003 (2011)
  doi:10.1088/0264-9381/28/19/195003
  [arXiv:1104.4199 [gr-qc]].



\bibitem{Gao:2018acg}
  Y.~X.~Gao, Y.~Huang and D.~J.~Liu,
  Phys.\ Rev.\ D {\bf 99}, no. 4, 044020 (2019)
  doi:10.1103/PhysRevD.99.044020
  [arXiv:1808.01433 [gr-qc]].

\bibitem{Zhang:2020pko}
  S.~J.~Zhang, B.~Wang, A.~Wang and J.~F.~Saavedra,
  Phys.\ Rev.\ D {\bf 102}, no. 12, 124056 (2020)
  doi:10.1103/PhysRevD.102.124056
  [arXiv:2010.05092 [gr-qc]].






\end{thebibliography}
\end{document}